\documentclass[%
 aip,
 jcp,%
 amsmath,amssymb,
 preprint,%
]{revtex4-1}

\usepackage{graphicx}% Include figure files
\usepackage{dcolumn}% Align table columns on decimal point
\usepackage{bm}% bold math
\usepackage[mathlines]{lineno}% Enable numbering of text and display math
\begin{document}

\preprint{AIP/123-QED}

\title[Trimer Self-Assembly and Fluid Phase Behavior]{Computational Study of Trimer Self-Assembly and Fluid Phase Behavior\footnote{Contribution of the National Institute of Standards and Technology, not subject to U.S. Copyright.}}% Force line breaks with \\
%\thanks{Contribution of the National Institute of Standards and Technology, not subject to U.S. Copyright.}

\author{Harold W. Hatch}
 \email{harold.hatch@nist.gov}
\affiliation{ 
Chemical Informatics Research Group, Chemical Sciences Division, National Institute of Standards and Technology, Gaithersburg, Maryland 20899-8380, USA
}%
\author{Jeetain Mittal}%
 \email{jeetain@lehigh.edu}
\affiliation{ 
Department of Chemical and Biomolecular Engineering, Lehigh University, Bethlehem, Pennsylvania 18015, USA
}%

\author{Vincent K. Shen}
% \email{vincent.shen@nist.gov}
\affiliation{ 
Chemical Informatics Research Group, Chemical Sciences Division, National Institute of Standards and Technology, Gaithersburg, Maryland 20899-8380, USA
}%
%\affiliation{%
%%Second institution and/or address%\\This line break forced% with \\
%Chemical Informatics Research Group, Chemical Sciences Division, National Institute of Standards and Technology, Gaithersburg, Maryland 20899-8380, USA
%}%

\date{\today}% It is always \today, today,
             %  but any date may be explicitly specified

\begin{abstract}
The fluid phase diagram of trimer particles composed of one central attractive bead and two repulsive beads was determined as a function of simple geometric parameters using flat-histogram Monte Carlo methods.
A variety of self-assembled structures were obtained including spherical micelle-like clusters, elongated clusters and densely packed cylinders, depending on both the state conditions and shape of the trimer.
Advanced simulation techniques were employed to determine transitions between self-assembled structures and macroscopic phases using thermodynamic and structural definitions.
Simple changes in particle geometry yield dramatic changes in phase behavior, ranging from macroscopic fluid phase separation to molecular-scale self-assembly.
In special cases, both self-assembled, elongated clusters and bulk fluid phase separation occur simultaneously.
Our work suggests that tuning particle shape and interactions can yield superstructures with controlled architecture. 
\end{abstract}

%Use showkeys class option if keyword
                              %display desired
\maketitle

%***********************************************************
\section{Introduction}
%***********************************************************

Biological molecules self-assemble into membranes, protein assemblies, viruses and cells.\cite{whitesides_self-assembly_2002}
%\textbf{(NOTE TO JM: help improve, more examples and refs?)}.
Material design inspired by nature is a promising route to create materials with novel or enhanced properties by spontaneous self-assembly.\cite{zheng_fabrication_2005, ozin_nanofabrication_2009, soukoulis_past_2011, vignolini_3d_2012}
In the laboratory, colloidal particles can be synthesized with a variety of shapes and directional interactions.\cite{sacanna_engineering_2013}
These patchy particles could potentially be used to mimic the self-assembly observed at smaller length scales, and to rationally design assemblies from their basic building blocks.\cite{glotzer_anisotropy_2007}

Studies of self-assembly range from those considering only repulsive interactions which define the shape of the particle,\cite{damasceno_predictive_2012, anders_understanding_2014} to those considering spherical particles with directional attractions.\cite{zhang_self-assembly_2004, chen_directed_2011, romano_colloidal_2011}
In colloidal systems, both the shape and the directional interactions are intimately coupled when depletant is added to the solution.\cite{cademartiri_using_2012}
This depletant interaction drives the assembly of lock-and-key colloids.\cite{sacanna_lock_2010}
For colloidal clusters synthesized with smooth and rough beads, the smooth beads attract more strongly to one another than to rough beads, due to more excluded volume overlap at contact.\cite{kraft_surface_2012}
The focus of this paper is on the self-assembly of trimers, consisting of a central attractive bead and two repulsive beads.

In an experimental and computational study, it was observed that dimers with one attractive bead and one repulsive bead self-assembled into spherical micelles.\cite{kraft_surface_2012}
In addition, it was demonstrated that trimers with one attractive bead and two repulsive beads could be synthesized.
Recently, an experimental and computational study of trimers with one attractive bead and two repulsive beads reported that only elongated clusters were formed, in contrast to the spherical micelles formed by dimers.\cite{wolters_self-assembly_2015} 
In a different computational study, a flexible 3-mer chain with two attractive beads and one repulsive bead at the end of the chain was found to self-assemble.\cite{jimenez-serratos_monte_2013}
Tetramers with two attractive beads and two repulsive beads were found to self-assemble into a variety of structures and used to study protein aggregation.\cite{barz_minimal_2014}

In this computational study, the phase behavior of a family of trimer models with one attractive central bead and two repulsive beads is investigated for a range of different trimer shapes.
Advanced simulation methods were used to obtain the fluid phase behavior based upon thermodynamic and structural definitions, rather than more phenomenological approaches.
In particular, Wang-Landau Transition Matrix Monte Carlo (WL-TMMC) simulations were preformed in the grand canonical ensemble, utilizing on the order of hundreds of billions of trials per simulation.
The trimers form spherical micelle-like clusters, elongated clusters and densely packed cylinders.
We show that there is a transition from self-assembly to bulk fluid phase separation as bond length is reduced, and find the in-between bond length with both elongated, self-assembled structures and fluid phase separation.
We also discuss how the phase behavior of the family of trimer models may be understood in terms of the interaction between particles.

This paper is organized as follows.
In Sec. \ref{sec:models}, we describe the family of trimer models studied in this work.
We then discuss the computational methods and the thermodynamic and structural transition definitions in Sec. \ref{sec:methods}.
Results are discussed in Sec. \ref{sec:results}.
Finally, we conclude and discuss future work in Sec. \ref{sec:conclusions}.

%***********************************************************
\section{\label{sec:models}Models}
%***********************************************************

\begin{figure}
\begin{center}
\includegraphics[width=8.5cm]{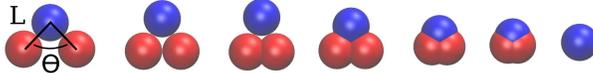}
\caption{
The family of trimer models investigated in this work illustrated using VMD.\cite{humphrey_vmd:_1996}
Blue beads are attracted to other blue beads, while all other pair interactions (red-red and blue-red) are purely repulsive.
The trimers are listed in order of increasing attractive region with respect to repulsive region, and the same order as Table \ref{tab:modelparams}.
}
\label{fig:trimermod}
\end{center}
\end{figure}

In this paper, we studied the fluid phase behavior of a family of trimer models.
The trimer consisted of one central, attractive bead and two repulsive beads, as shown in Figure \ref{fig:trimermod} and Table \ref{tab:modelparams}.
Specifically, we studied how fluid phase behavior was affected by simple geometric parameters, $L$ and $\Theta$, where $L$ is the rigid bond length between an attractive and repulsive bead, and $\Theta$ is the rigid bond angle with the vertex on the attractive bead.
%These are shown in Figure \ref{fig:trimermod} and Table \ref{tab:modelparams}.
The beads interact via a shifted-force Lennard-Jones (LJ) potential,
\begin{eqnarray}
U_{LJ}^{SF}(r) = U_{LJ}(r) - U_{LJ}(r_c) - (r-r_c) \left. \frac{\partial U_{LJ}}{\partial r} \right|_{r=r_c} 
\\
U_{LJ}(r) = 4 \epsilon \left[ \left(\frac{\sigma}{r}\right)^{12} - \left(\frac{\sigma}{r}\right)^6 \right]
\end{eqnarray}
where $r_c$ is the potential cut-off, $U(r \geq r_c)=0$.
For interactions between attractive, blue beads, $r_c/\sigma=3$.
All other pair-wise interactions are purely repulsive, $r_c/\sigma=2^{1/6}$, also known as the Weeks-Chandler-Andersen potential.\cite{weeks_role_1971}
Each bead has equal $\sigma, \epsilon$ and mass.

\begin{table}
\caption{\label{tab:modelparams}
Trimer model parameters, $L$ and $\Theta$, and computed values for the excluded volume (see Appendix \ref{sec:exvol}), critical temperature, and Boyle temperature (see Appendix \ref{sec:boyle}).
}
\begin{ruledtabular}
\begin{tabular}{cc|cccc}
 $L/\sigma$ & $\Theta$ & $V_{ex}/\sigma^3 $ & $k_B T_\textrm{c}/\epsilon$ & $k_B T_{\textrm{Boyle}} / \epsilon$ \\
\hline
 1    & $\pi/2$ & 9.83   & n/a   & 0.365(5)\\
 1    & $\pi/3$ & 9.31   & n/a   & 0.435(5)  \\
 1    & $\pi/4$ & 8.88   & n/a   & 0.485(5)  \\
 0.75 & $\pi/3$ & 8.02   & n/a   & 0.505(5)  \\
 0.4  & $\pi/3$ & 6.19   & 0.3117(1)  & 0.815(5)  \\
 0.25 & $\pi/3$ & 5.41   & 0.4989(1)  & 1.17(1)  \\
 0    & $\pi/3$ & 4.19   & 0.8798(7)  & 2.00(2)  \\
\end{tabular}
\end{ruledtabular}
\end{table}

\begin{figure}
\centering
\includegraphics[width=8.5cm]{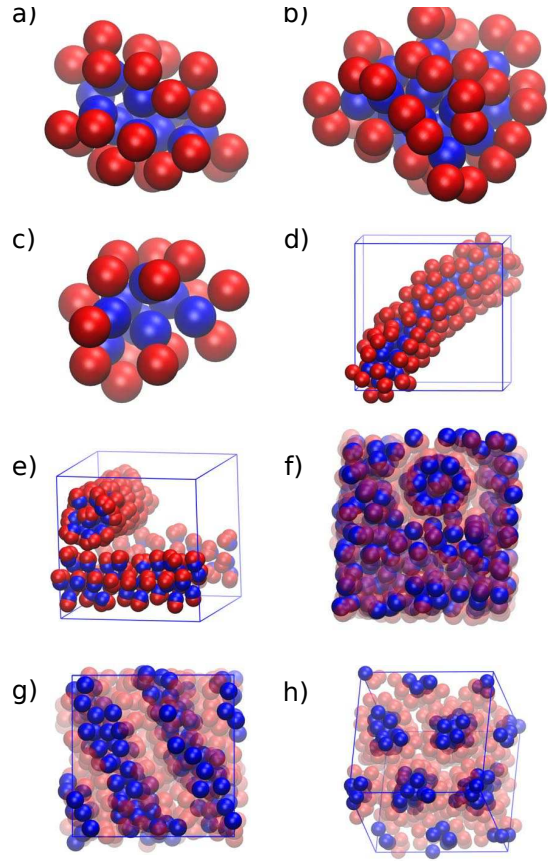}
%\subcaptionbox{\label{fig:micelle60_1}}
%%{\includegraphics[width=4cm]{fig_2a.eps}}
%{\includegraphics[width=4cm]{../pictures/micelle60_1/vmdscene.eps}}
%\subcaptionbox{\label{fig:micell45_1}}
%%{\includegraphics[width=4.5cm]{fig_2b.eps}}
%{\includegraphics[width=4.5cm]{../pictures/micelle45_1/vmdscene.eps}}
%\subcaptionbox{\label{fig:micelle90_1}}
%%{\includegraphics[width=3.5cm]{fig_2c.eps}}
%{\includegraphics[width=3.5cm]{../pictures/micelle90_1/vmdscene.eps}}
%\subcaptionbox{\label{fig:string60_1}}
%%{\includegraphics[width=4cm]{fig_2d.eps}}
%{\includegraphics[width=4cm]{../pictures/string60_1/vmdscene.eps}}
%\subcaptionbox{\label{fig:string60_0.4}}
%%{\includegraphics[width=4cm]{fig_2e.eps}}
%{\includegraphics[width=4cm]{../pictures/string60_0.4/vmdscene.eps}}
%\subcaptionbox{\label{fig:logliquid}}
%%{\includegraphics[width=4cm]{fig_2f.eps}}
%{\includegraphics[width=4cm]{../pictures/logliquid/vmdscene.eps}}
%\subcaptionbox{\label{fig:cylinder60_1}}
%%{\includegraphics[width=4cm]{fig_2g.eps}}
%{\includegraphics[width=4cm]{../pictures/cylinder60_1/vmdsceneside.eps}}
%\subcaptionbox{\label{fig:cylinder60_1_2}}
%%{\includegraphics[width=4cm]{fig_2h.eps}}
%{\includegraphics[width=4cm]{../pictures/cylinder60_1/vmdscenetop.eps}}
%\subcaptionbox{\label{fig:liquid60_0.4}}
%%{\includegraphics[width=4cm]{fig_2i.eps}}
%{\includegraphics[width=4cm]{../pictures/liquid60_0.4/vmdscene.eps}}
\caption{Illustration of selected structures.
Unless otherwise specified, $L=\sigma$, $\Theta=\pi/3$, $k_BT/\epsilon=0.2$, $V=729\sigma^3$ and the blue boxes represent periodic boundaries. 
(a) micelle, $N=13$
(b) large micelle, $\Theta=\pi/4$, $N=20$
(c) small micelle, $\Theta=\pi/2$, $N=8$
(d) elongated cluster, $k_BT/\epsilon=0.125$, $N=59$ 
(e) elongated cluster, $L=0.4\sigma$, $k_BT/\epsilon=0.15$, $N=112$ 
(f) elongated cluster in liquid, $L=0.4\sigma$, $k_BT/\epsilon=0.15$, $\rho V_{ex}=2.6$ 
(g) cylinder, $\rho V_{ex}=2.2$ 
(h) same as (g) with a top-down projection
%(i) liquid, $L=0.4\sigma$, $k_BT/\epsilon=0.25$, $\rho V_{ex}=2.6$.
}\label{fig:saPictures}
\end{figure}

While the model described above may seem simplistic, it is intended to capture basic geometric features that should be relevant to a broad range of systems.
Indeed, this trimer model exhibited rich phase behavior with respect to self-assembly and fluid phase separation (see Figures \ref{fig:saPictures} and \ref{fig:vle}).
The aim of this study is to rationalize how the phase behavior and self-assembly changes with particle shape, using a general model that may be applied to many different types of systems and is computationally tractable.
In this study, the Lennard-Jones potential was chosen for simplicity.

%***********************************************************
\section{\label{sec:methods}Methods}
%***********************************************************

Flat-histogram sampling methods were used to investigate the fluid phase behavior of the family of trimer models.
Specifically, Wang-Landau Transition-Matrix Monte Carlo (WL-TMMC) simulations\cite{errington_direct_2003, shell_improved_2003, rane_monte_2013} in the grand canonical ensemble were performed, as described below in Section \ref{sec:methodswltmmc}.
This powerful simulation method computes the free energy, potential energy and pressure as a function of density at constant temperature, as well as provides detailed structural information, in a single simulation.
The advantage of the grand-canonical ensemble over the canonical ensemble is that smaller system volumes can be used to capture physically relevant density fluctuations.
In the canonical ensemble, where the total number of particles is fixed, the use of small system volumes amounts to the imposition of a constraint.\cite{corti_statistical_1998}
For self-assembling systems, this arbitrary constraint may not agree with the preferred free monomer densities and sizes of self-assembled structures in the thermodynamic limit.
The effect of this constraint in the canonical ensemble diminishes with system size, and thus appropriate canonical-ensemble simulations of self-assembly require significantly larger systems that are computationally more expensive.
In addition, to improve sampling at low temperature, WL-TMMC was combined with parallel-tempering.
A single isotherm simulation was typically composed of hundreds of billions of trials.
Simulation details are provided in Section \ref{sec:methodswltmmc}, and the methods used to determine phase coexistence and locate structural transitions are described in Section \ref{sec:methodspb}.

%***********************************************************
\subsection{\label{sec:methodswltmmc}Grand Canonical Wang-Landau Transition-Matrix Monte Carlo}
%***********************************************************

The Grand canonical WL-TMMC method was used to obtain the macrostate probability distribution, $\Pi(N;\mu,V,T)$, which is the probability to observe the number of trimers, $N$, for a given chemical potential, $\mu$, volume, $V$, and temperature, $T$.
See Appendix A of Ref. \onlinecite{shen_elucidating_2014} for implementation details of WL-TMMC used here.
The Wang-Landau update factor was initially set to unity, and was multiplied by 0.5 whenever the flatness criteria of 80 \% was met.
After the update factor was smaller than $10^{-6}$, the collection matrix was updated.
After the update factor was smaller than $5 \times 10^{-8}$, transition-matrix Monte Carlo was performed with an update to the biasing function every $10^6$ trials.
To parallelize the single isotherm simulations, each isotherm was divided into 12 overlapping windows to span the entire density range, that is the range of $N$ values from $0$ to $N_{max}$.
Because lower density simulations are faster, the parallelization was load balanced by decreasing window size with increasing density by a power scaling with exponent of 1.5.
Neighboring windows overlapped by up to three trimers.
The free energy of the entire density range was recovered by setting the free energy of neighboring windows equal at the middle overlapping trimer number, and discarding the largest and smallest number of trimers (when neighbor present) in each window.
A window was converged if it swept at least 10 times, although most windows swept 100-1000 times while waiting for the high density window to converge.
A sweep was defined as satisfying the condition that each macrostate had been visited from a different macrostate at least 100 times.
After a simulation swept more than one time, canonical ensemble averages, as described below, were accumulated for quantities such as the potential energy and the squared potential energy upon every successful trial attempt.
Ideally, when running a WL-TMMC simulation, a value of the chemical potential is chosen such that the difference between neighboring macrostates in the macrostate distribution is minimized.
Fortunately, the exact choice of the $\mu$ in the WL-TMMC simulations is relatively unimportant, because the initial Wang-Landau part of the simulation efficiently finds the order of magnitude of the macrostate distribution, and then the macrostate distribution may be histogram reweighted to different values of $\mu$.

The following Monte Carlo trials were employed, as listed in Table \ref{tab:mcmoves}.
Rigid trimer translations or rotations about the center of mass were attempted with equal probability.
Random insertions or deletions of trimers were also attempted, subject to Metropolis acceptance criteria.\cite{frenkel_understanding_2002}
Collective trial moves were also implemented to facilitate convergence with self-assembled structures.
Smart Monte Carlo was used to bias the movement of trimers in the direction of their center-of-mass forces.\cite{rossky_brownian_1978}
A second collective move type entailed rigid translation or rotation of each cluster of trimers.
Clusters were defined as all trimers having an attractive bead within a cut-off distance, $4\sigma/3$, from at least one other attractive bead in the cluster, obtained via recursive flood-fill algorithm.
To obey detailed balance, cluster moves which resulted in a trimer joining a different cluster were rejected.
Statistics on the clusters were accumulated every attempted cluster move, after the simulation swept more than one time.
For each Monte Carlo trial that involved movement of trimers, the parameter associated with the maximum possible translation or rotation was optimized, via a 5 \% change every $10^6$ trials, to yield approximately 25 \% acceptance of the trial move.
Another Monte Carlo trial involved configuration swaps between neighboring density windows to facilitate convergence.
Configurational swap moves between adjacent density windows were used to ensure self-assembled structures were sampled in multiple windows.
These configurational swap moves helped to improve the parallelization efficiency, and were performed at fixed $N, V, T$, and $\mu$.\cite{frenkel_understanding_2002}
Even with the assortment of trial moves described above, structural transitions between different self-assembled motifs were difficult to sample at low $T$.
To circumvent this difficulty, parallel tempering was implemented to swap configurations between neighboring temperatures, at fixed number of trimers, from a series of closely spaced isotherms.
The second type of configurational swap move improved sampling of structural transitions that occurred as temperature is decreased, and was performed at fixed $N, V$ with varying $U, T, \mu$.
When the configuration swap trial is attempted, there is a 50 \% chance to store the current configuration, and a 50 \% chance to swap the current configuration with a stored configuration on an overlapping processor (if exists), subject to Metropolis acceptance criteria.\cite{frenkel_understanding_2002}

\begin{table}
\caption{\label{tab:mcmoves}
Monte Carlo trials and weights for the probability of selection.
}
\begin{ruledtabular}
\begin{tabular}{llc}
 trial & weight \\
\hline
 single-trimer translation or rotation   &  1       \\
 single-trimer insertion or deletion     &  1/4    \\
 smart Monte Carlo\cite{rossky_brownian_1978}    & $1/10N_{max}$   \\
 cluster translation or rotation  & $1/5N_{max}$   \\
 parallel configuration swap   & $5 \times 10^{-6}$   \\
\end{tabular}
\end{ruledtabular}
\end{table}

Grand canonical ensemble averages of an observable, $A$, denoted as $\langle A \rangle_{\mu VT}$, are obtained as a continuous function of $\langle N \rangle_{\mu VT}$.
Calculation of $\langle A\rangle_{\mu VT}$ is based on the canonical average of property $A$, denoted as $\langle A\rangle_{NVT}$, which can be calculated during the course of the simulation,
\begin{equation}
\langle A(N)\rangle_{NVT} = \frac{\sum_{i=0}^{N_{trial}} A(i) \delta(n_i - N) }{ \sum_{i=0}^{N_{trial}} \delta(n_i - N)}
\end{equation}
where $n_i$ is the number of trimers in trial $i$, $\delta$ is the delta function, and $N_{trial}$ are the number of sampled states in the simulation.
It follows that the grand-canonical average is
\begin{equation}
\langle A(\langle N\rangle_{\mu VT})\rangle_{\mu VT} = \sum_{n=0}^{N_{max}} \langle A(n)\rangle_{NVT} \Pi(n;\mu).
\label{eq:gcensav}
\end{equation}
The quantity $\langle N\rangle_{\mu VT}$ can be obtained directly from the macrostate distribution as
\begin{equation}
\langle N\rangle_{\mu VT} = \sum_{n=0}^{N_{max}} n \Pi(n;\mu).
\end{equation}
The average properties at other state conditions, namely different values of $\mu$, can be obtained via histogram reweighting the macrostate distribution.

%***********************************************************
\subsection{\label{sec:methodspb}Determining Phase Coexistence and Structural Transitions}
%***********************************************************

In this section, we discuss the methods used to determine fluid phase behavior.
The two distinct types of behavior observed in this work are macroscopic phase separation and self-assembly (e.g. micellization), which also includes transitions between different structures. 
Note that structural transitions that take place on a microscopic length scale, such as micellization, are not true thermodynamic phase transitions.\cite{floriano_micellization_1999}
Phase coexistence conditions between two macroscopic phases were obtained by histogram reweighting the macrostate distribution to a value of $\mu$ such that the probabilities of observing each phase are equal.
Critical points were obtained by fitting saturation densities to the law of rectilinear diameters.\cite{frenkel_understanding_2002}

For the remainder of this section, we discuss the methods used to determine the structural transitions involving the spherical micellar fluid, which requires locating the low and high density and low and high temperature boundaries for the spherical micellar fluid.
The low density boundary is defined primarily by the critical micelle concentration, which is the concentration above which a free trimer fluid becomes a micellar fluid.
Similarly, the high temperature boundary is taken to be the maximum temperature at which micelles are stable.
At low temperature, the micellar fluid transforms from roughly spherical structures to elongated ones.
The temperature at which this occurs is taken to be the low-temperature boundary of the spherical micellar fluid.
The high density transition of micelles into more solid-like structures is also approximately obtained, although proper sampling at high densities is beyond the scope of this work.
Examples of these transitions are provided in Appendix \ref{app:saexample}.

The critical micelle concentration (CMC), defined as the lowest concentration at which micelles can form, was obtained by both thermodynamic and structural definitions.
The structural method directly measures the CMC as the concentration of free trimers and premicellar aggregates as a function of density.\cite{lebard_self-assembly_2012}
This direct measurement of the CMC is possible because the concentration of free trimers and premicellar aggregates, $\rho_{free}$, remains approximately constant as the fluid density, $\langle N\rangle/V$, increases at fixed temperature after micelle formation.\cite{floriano_micellization_1999}
Premicellar aggregates are defined as clusters with a number of trimers less than or equal to the first minimum in the histogram of aggregate size (typically 4-5 trimers), and clusters are defined in Section \ref{sec:methodswltmmc}.
$\langle\rho_{free}\rangle_{\mu VT}$ is relatively constant over a fluid density range.
In practice, the density range over which $\rho_{free}$ is constant was defined as the range where $\rho_{free}$ was within some tolerance of the first local maximum.
In this work, we used a 75 \% tolerance.
The structurally based critical micelle concentration was taken as the average $\rho_{free}$ in this fluid density range, and the high density boundary of the micellar fluid was taken as the maximum density in this fluid density range.
The thermodynamic method to obtain the CMC uses the density at which the equation of state deviates from ideal trimer fluid behavior.\cite{panagiotopoulos_micellization_2002}
The deviation appears as a second linear regime, due to the formation of micelles, and the density where the deviation occurs is defined by the point of intersection of fits to the linear regimes.
The equation of state is obtained from the macrostate distribution, by reweighting it to various $\mu$ values, and computing the pressure as a function of $\langle N\rangle_{\mu VT}/V$ by comparing the probability to observe zero trimers to the ideal trimer fluid state.\cite{errington_direct_2003}
In order to precisely obtain the equation of state in the density range of interest with WL-TMMC simulations in the grand canonical ensemble, both $V$ and $N_{max}$ must be tuned to sample both the ideal and micellar fluids.
Depending on the temperature, $V/\sigma^3$ ranged from $9^3$ to $64^3$ while $N_{max}$ was 50 to 150.
These low density simulations for the thermodynamic definition of the CMC were separate from the higher density simulations which were used to obtain the CMC by the structural definition.

The critical micelle temperature (CMT) was taken to be the highest temperature at which micelles could exist.
This temperature is not a true critical point, and was simply named by analogy to the critical micelle concentration.\cite{floriano_micellization_1999}
As noted in previous work, defining the CMT is somewhat arbitrary.\cite{floriano_micellization_1999}
Although one may define the CMT with structural information, it is difficult to distinguish between self-assembled micelles and supercritical clusters, similar to those formed in typical homogeneous fluids.
In previous work, a thermodynamic signature of micellization was a system-size dependent density of a second peak in the macrostate probability distribution of the number of particles.\cite{salaniwal_competing_2003}
Physically, the second peak corresponds to the formation of a micelle, which happens at the same number of trimers, regardless if the system size is slightly increased.
This thermodynamic signature of micellization was used to define the CMT in this work, as demonstrated in Appendix \ref{app:saexample}.
The error bars for the CMT simply depended on the spacing between simulated isotherms.
Specifically, the CMT was obtained by identifying two isotherms.
The first is the highest temperature in which the fluid contained micelles, and the second is the next highest temperature in which the fluid did not contain micelles.
This effectively brackets the CMT.
Therefore, the CMT must be in between these two isotherms.
The reported value of the CMT was the average of these two isotherms, and the size of the error bar in temperature is half of the difference in temperature of these two isotherms.
Finally, the density associated with the CMT is the critical micelle concentration at the CMT.
A conservative estimate of the CMC at the CMT was obtained from the highest temperature simulated isotherm which contained micelles.
Using this isotherm, the CMC at the CMT is within the density range between the CMC and high density boundary of the micellar fluid just below the CMT.

Structural transitions at low temperature were determined using parallel tempering simulations.
Conventional WL-TMMC simulations in the grand canonical ensemble at fixed $T$ roughly identified the temperature region where elongated clusters formed, and where to perform the parallel tempering simulations.
In the parallel tempering simulations, many isotherms from grand canonical WL-TMMC simulations were performed at closely spaced intervals in temperature to target the micelle-to-elongated cluster transition region, and the isotherms were allowed to exchange configurations between temperatures at constant number of trimers.
The configuration exchanges in parallel tempering allowed for more frequent sampling of the micelle to elongated cluster transition.
This transition was identified by both structural and thermodynamic definitions.
In the structural definition, the transition occurs at the temperature at which there is equal probability of observing more than one micelle, and one elongated cluster.
The maximum number of trimers, $N_{max}$, in each isotherm was set to a value which would typically contain two micelles, when above the transition temperature.
The thermodynamic definition is the temperature at which there is a peak in the constant volume heat capacity.\cite{bhattacharya_critical_2001}
The constant volume heat capacity, $C_V$ is computed as
\begin{equation}
\langle C_V\rangle_{NVT} = \frac{\langle U^2\rangle_{NVT}-\langle U\rangle_{NVT}^2}{k_BT^2}
\end{equation}
where $U$ is the potential energy and $\langle ...\rangle_{NVT}$ is a canonical ensemble average.
The grand canonical ensemble average is then obtained from Equation \ref{eq:gcensav}.
In these parallel tempering simulations, twelve isotherms were simulated every $\Delta \frac{\epsilon}{k_B T}=0.25$, in the range $\frac{\epsilon}{k_B T} \in [4.5, 7.25]$.
In order to study the density dependence of this transition temperature, two sets of parallel tempering simulations were performed with different volumes, $V/\sigma^3=729$ and $5832$.

%***********************************************************
\section{\label{sec:results}Results and Discussion}
%***********************************************************

We studied the phase behavior of a trimer fluid as a function of bond length $L$ and bond angle $\Theta$.
The interesting finding is that dramatic changes in phase behavior can be caused by simple changes in the geometry of the trimer.
In particular, the phase behavior changes dramatically from macroscopic fluid phase separation without self-assembly at low $L$ to self-assembly without fluid phase separation as $L$ increases up to $L=\sigma$.
In the special in-between case of $L=0.4\sigma$, both fluid phase separation and self-assembly occurred simultaneously, where the latter resulted in the formation of elongated clusters.

A variety of self-assembled structures were observed for $L/\sigma = 0.4, 0.75, 1$, as shown in Figure \ref{fig:saPictures}.
In particular, two predominant types of self-assembled structures formed in the density range of interest in this study.
The first type of self-assembled structures can be described as micelle-like spherical clusters.
These micelles were of variable size, depending on both the state conditions and the shape parameters of the trimer model, shown in Figures \ref{fig:saPictures}a, \ref{fig:saPictures}b, and \ref{fig:saPictures}c.
The second type of self-assembled structure can be described as elongated clusters, shown in Figures \ref{fig:saPictures}d, \ref{fig:saPictures}e, and \ref{fig:saPictures}f.
One important feature of elongated clusters is that they may form at low density.
Also note that these elongated clusters may vary in shape, depending on both $L$ and $\Theta$.
A third type of self-assembled structure, packed cylinders, shown in Figures \ref{fig:saPictures}g and \ref{fig:saPictures}h, is only observed at high density.
Simulations in this high density regime are beyond the scope of this study due to sampling difficulties.

\begin{figure}
\begin{center}
\includegraphics[width=8.5cm]{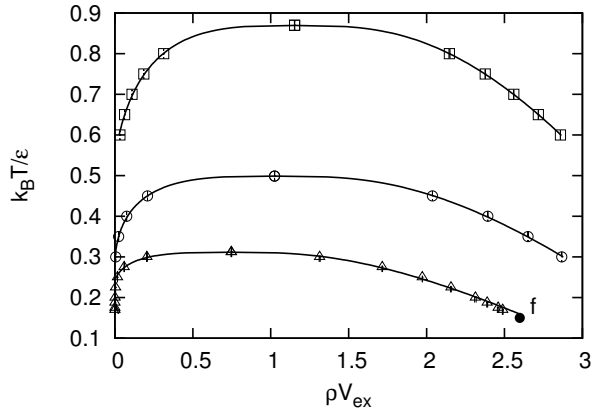}
\caption{
Fluid phase coexistence for (squares) $L=0$, (circles) $L=0.25\sigma$, (triangles) $L=0.4\sigma$ with $\Theta=\pi/3$.
The critical points, shown by the symbols at the maximum temperature, were obtained from fit to the law of rectilinear diameters.
The labeled black circle corresponds with the structure shown in Figure \ref{fig:saPictures}f.
The error bars, smaller than symbols, were obtained from three independent simulations.
}\label{fig:vle}
\end{center}
\end{figure}

\begin{figure}
\begin{center}
\includegraphics[width=8.5cm]{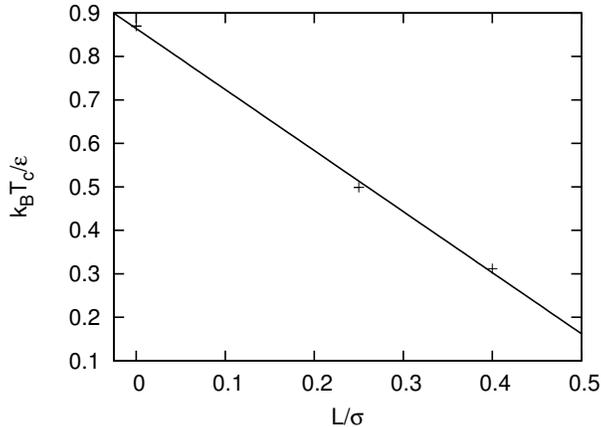}
\caption{
Critical temperature as a function of bond length with $\Theta=\pi/3$, shown by symbols, with linear fit.
The error bars, smaller than symbols, were obtained from three independent simulations.
}\label{fig:tcl}
\end{center}
\end{figure}

Fluid phase separation was observed for $L/\sigma = 0, 0.25, 0.4$ and $\Theta = \pi/3$, as shown in Figure \ref{fig:vle}.
In Figure \ref{fig:vle}, the density, $\rho$, is normalized by the excluded volume, $V_{ex}$ (see Table \ref{tab:modelparams}), in order for the phase coexistence curves for different values of $L$ to be in a similar density range.
For the largest bond length exhibiting fluid phase separation, $L=0.4\sigma$, both the low- and high-density coexisting liquids are inhomogeneous due to the presence of elongated, self-assembled structures (see Figures \ref{fig:saPictures}e and \ref{fig:saPictures}f).
Incidentally, we observe an approximately linear dependence of the critical temperature on the bond length in the range investigated, as shown in Figure \ref{fig:tcl}.
The line in Figure \ref{fig:tcl} is a linear fit, $k_BT_\textrm{c}^{fit}(L)/\epsilon=aL/\sigma+b$, where $a=-1.40(6)$ and $b = 0.86(2)$.
This linear trend is reminiscent of a linear trend in the critical temperature with respect to relative attraction between two beads of a dimer model reported previously.\cite{munao_phase_2014}

\begin{figure*}
\centering
\includegraphics[width=8cm]{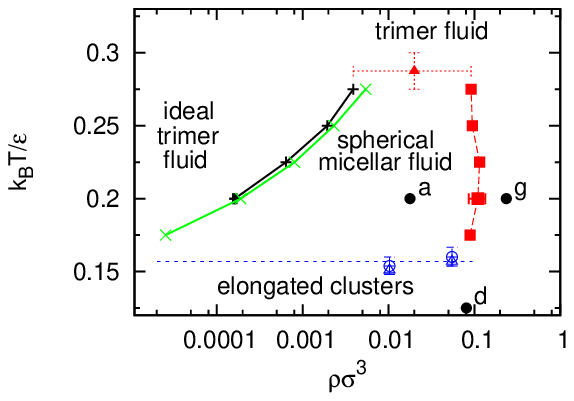}
\includegraphics[width=8cm]{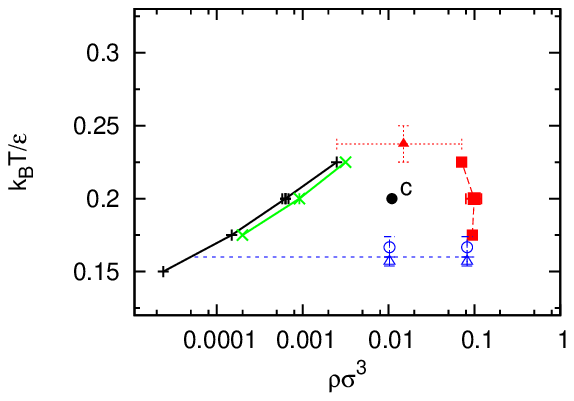}
\includegraphics[width=8cm]{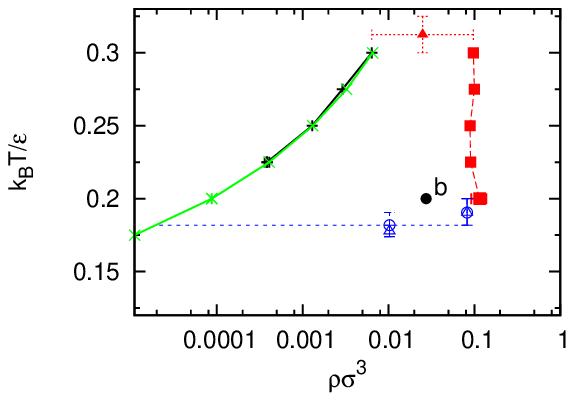}
\includegraphics[width=8cm]{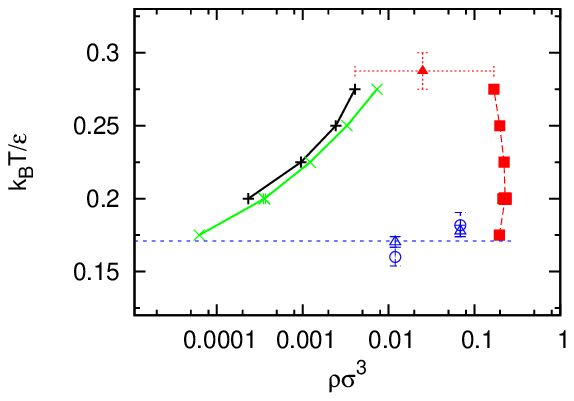}
\caption{Phase diagrams for (top left) $L=\sigma, \Theta=\pi/3$ (top right) $L=\sigma, \Theta=\pi/2$ (bottom left) $L=\sigma, \Theta=\pi/4$ (bottom right) $L=0.75\sigma, \Theta=\pi/3$.
The labeled black circles correspond with the structures shown in Figure \ref{fig:saPictures}.
The critical micelle concentration is shown with the (green x) structural definition and (black +) thermodynamic definition.
The critical micelle temperature is shown by the red triangle.
The micelle to elongated cluster transition is shown with the (open blue triangle) structural definition and (open blue square) thermodynamic definition.
The high density boundary of the micellar fluid is shown by the solid red squares.
Lines are guides to the eye.
}\label{fig:saPhaseDiagrams}
\end{figure*}

In contrast to the macroscopic phase separation observed for bond lengths $L \le 0.4\sigma$, trimer fluids with $L \ge 0.75\sigma$ self-assembled into micelles and did not exhibit macroscopic phase separation.
In Figure \ref{fig:saPhaseDiagrams}, we show phase diagrams using the approach described in Section \ref{sec:methodspb} and Appendix \ref{app:saexample} for the following four trimer model parameter pairs denoted as ($L$, $\Theta$): $(\sigma, \pi/3)$, $(\sigma, \pi/2)$, $(\sigma, \pi/4)$, $(0.75\sigma, \pi/3)$.
For all cases, the critical micelle concentration increased with temperature.
In addition, the critical micelle concentration increased with bond angle, $\Theta$, for $L=\sigma$ at fixed $T$.
At a select temperature, error bars for the CMC and the high density boundaries of the micellar fluid were computed as the standard deviation from three independent simulations with volumes, $V/\sigma^3=512,729,857.375$.
A secondary purpose for using different values of $V$ in the three independent simulations was to verify that the results were not system-size dependent, and indeed they were not dependent on system size.
The critical micelle temperature increased with decreasing bond angle, $\Theta$, for $L=\sigma$.
For the spherical to elongated cluster transition temperature, a set of parallel tempering simulations yielded a small density range for the transition temperature.
Recall that this represents a low-temperature boundary for the spherical micellar fluid.
In order to investigate the density dependence of this spherical to elongated cluster transition over a greater density range, two sets of parallel tempering simulations were performed at volumes $V/\sigma^3=729$ or $5832$.
The spherical to elongated cluster transition appeared to be relatively insensitive to density within the error bars of the simulations.

\begin{figure*}
\centering
%{\includegraphics[width=8cm]{../../cg3_60_1_1/struc/plotsurf.eps}}
{\includegraphics[width=8cm]{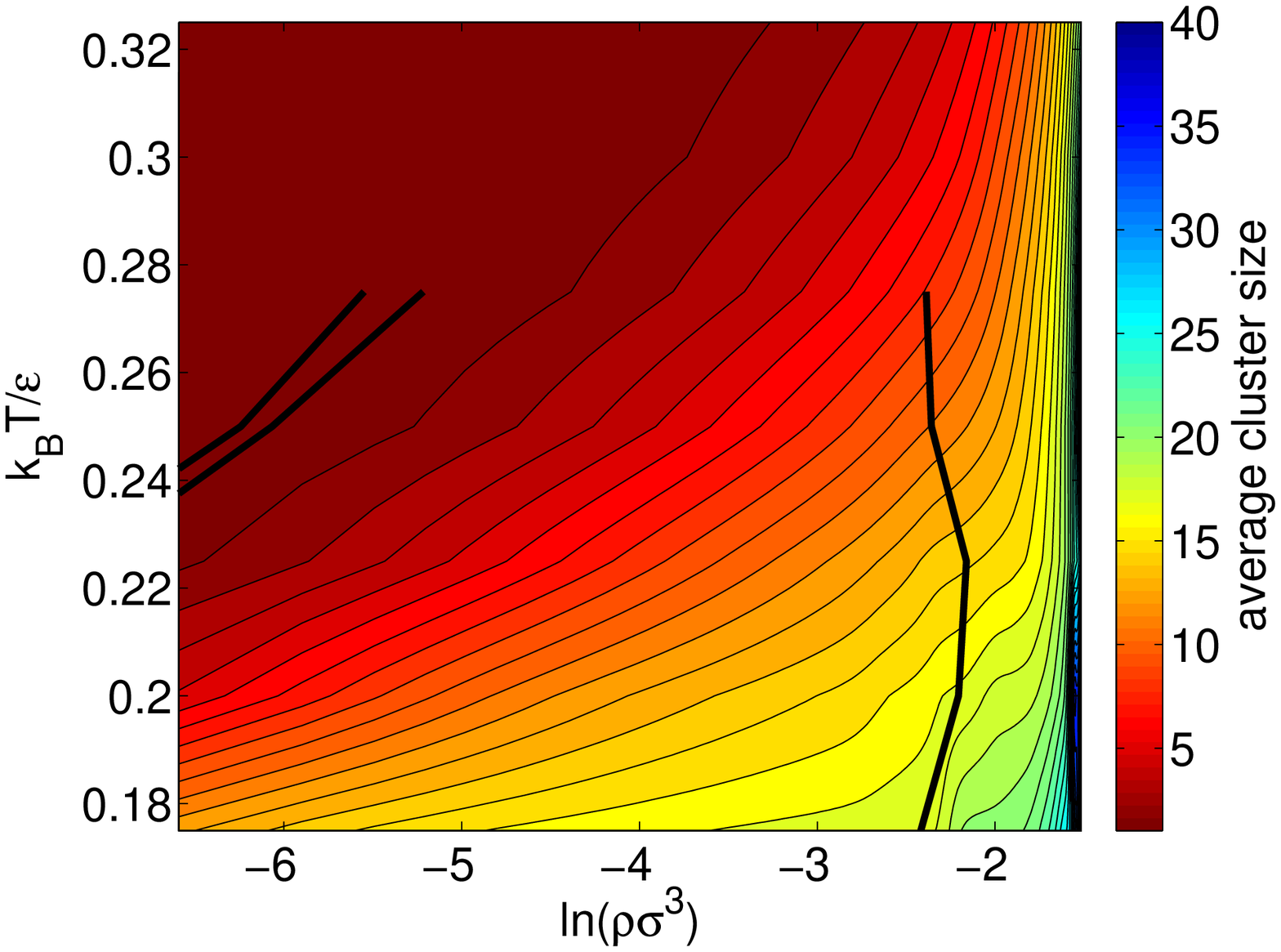}}
%{\includegraphics[width=8cm]{../../cg3_90_1_1/struc/plotsurf.eps}}
{\includegraphics[width=8cm]{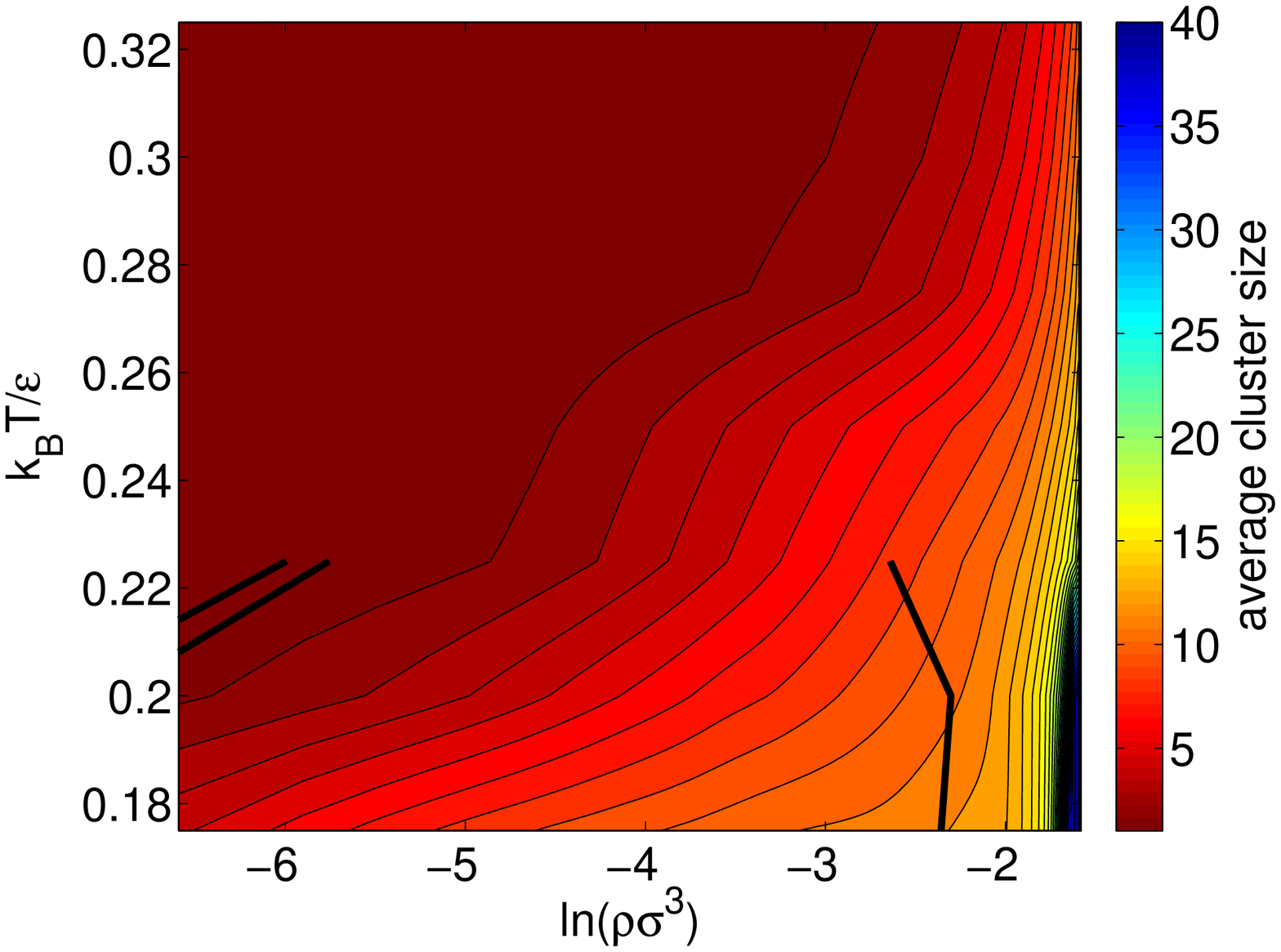}}
%{\includegraphics[width=8cm]{../../cg3_45_1_1/struc/plotsurf.eps}}
{\includegraphics[width=8cm]{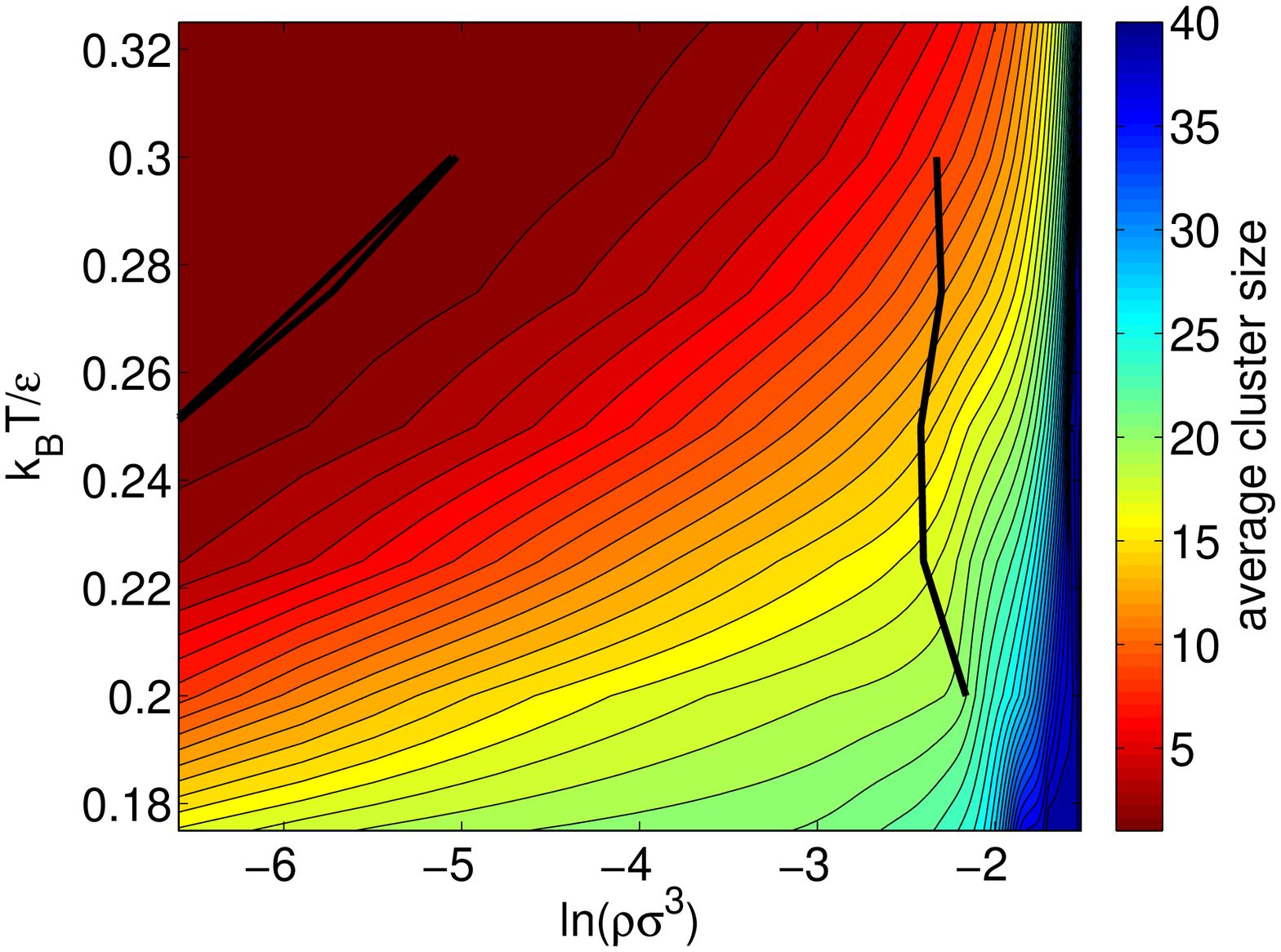}}
%{\includegraphics[width=8cm]{../../cg3_60_0.75_1/struc/plotsurf.eps}}
{\includegraphics[width=8cm]{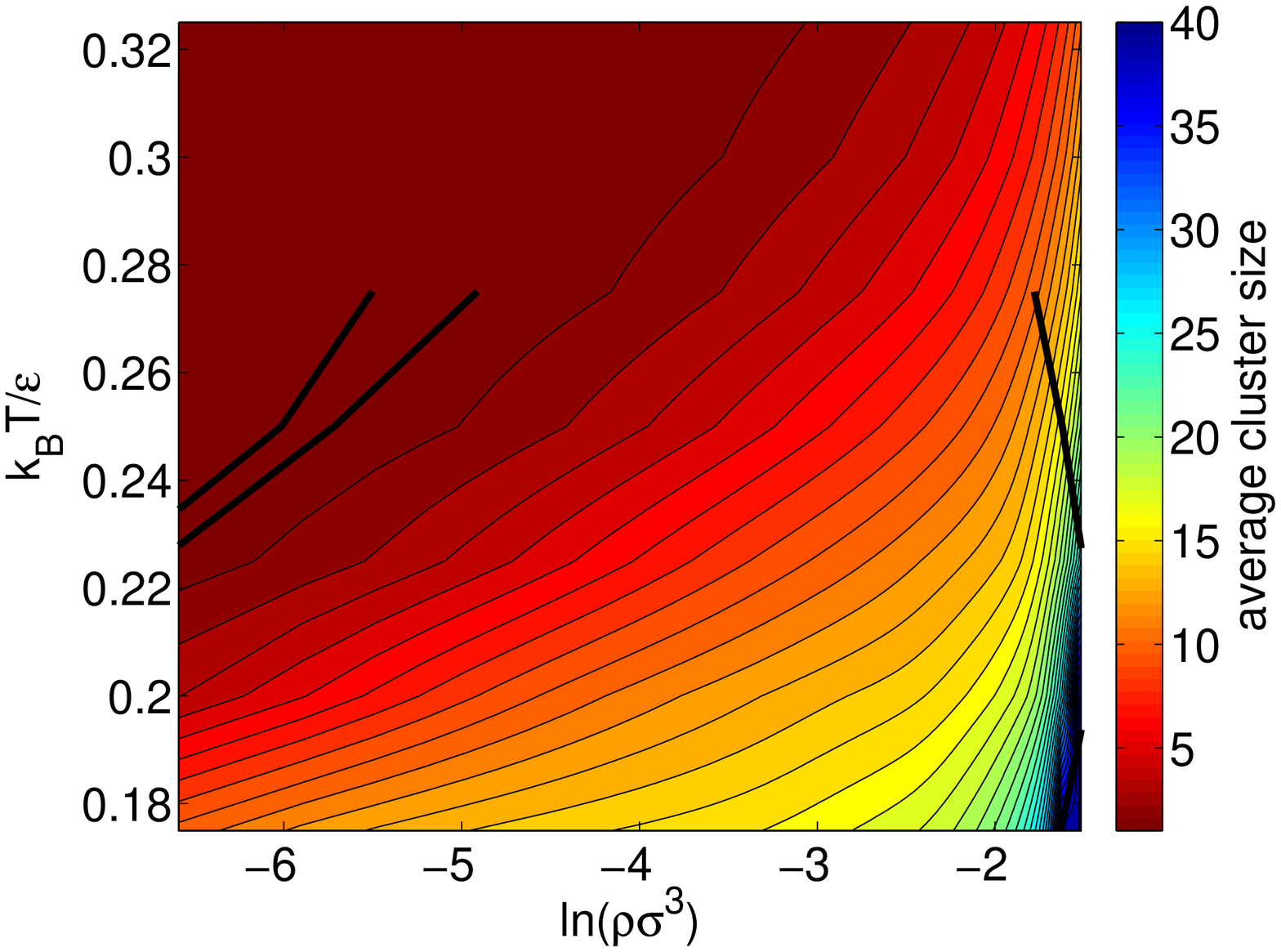}}
\caption{Average number of trimers in a cluster for (top left) $L=\sigma, \Theta=\pi/3$ (top right) $L=\sigma, \Theta=\pi/2$ (bottom left) $L=\sigma, \Theta=\pi/4$ (bottom right) $L=0.75\sigma, \Theta=\pi/3$.
Contour spacing is one trimer, and was obtained by 3-dimensional fit to a number of isotherms.
The thick, solid black lines are the boundaries shown in Figure \ref{fig:saPhaseDiagrams}.
The contours were truncated at 40 trimers.
}\label{fig:clusterMap}
\end{figure*}

In Figure \ref{fig:clusterMap}, the average cluster size as a function of temperature and density is shown for the following four trimer model parameter pairs denoted as ($L$, $\Theta$): $(\sigma, \pi/3)$, $(\sigma, \pi/2)$, $(\sigma, \pi/4)$, $(0.75\sigma, \pi/3)$.
Cluster sizes increase with increasing density and decrease with increasing temperature.
In addition, cluster sizes decrease as the bond angle, $\Theta$, is increased from $\pi/4$ to $\pi/2$.
Only temperatures above the spherical to elongated cluster transition are shown in Figure \ref{fig:clusterMap}.
This $T$ range was chosen because, when elongated clusters form, the majority of trimers in the system are part of a single cluster, and the cluster size is trivially equivalent to the number of trimers in the system.

The fluid phase behavior of the family of trimer models may be understood in terms of the relative size of the repulsive and attractive regions.\cite{salaniwal_competing_2003, munao_phase_2014, mahynski_grafted_2015}
As the repulsive region and anisotropy increases, self-assembly is more favored than macroscopic phase separation.
One way to quantify the relative sizes of the repulsive and attractive regions is to assume that the attractive region stays fixed while changes in the repulsive region are due to changes in the net excluded volume of the beads.
It thus follows that the smallest bond length corresponds to the smallest repulsive region, and the longest bond length corresponds to the largest repulsive region.
The models in Figure \ref{fig:trimermod} and Table \ref{tab:modelparams} are listed in order of decreasing excluded volume.
As the excluded volume increases with $L$, the trimer shifts from fluid phase separation at low $L$ to self-assembly at high $L$, with the special case of $L/\sigma=0.4$ possessing both fluid phase separation and self-assembly.
The relative change in the size of attractive to repulsive regions may also be observed via the Boyle temperature, $k_B T_{\textrm{Boyle}} / \epsilon$.
The Boyle temperature is the temperature at which the second virial coefficient is zero (see Section \ref{sec:methods}), where the attractive and repulsive contributions cancel to yield the compressibility of an ideal trimer fluid.
Finally, the relative size of the attractive region with respect to the repulsive region also explains the change in the critical temperature, $k_B T_\textrm{c} / \epsilon$, where $k_B$ is the Boltzmann constant.
Given that the trivial, isotropic case of $L=0$ possesses a critical temperature, increasing the repulsive region reduces the critical temperature.
Increasing the anisotropy of the model yielded self-assembled structures, but also reduced critical temperatures.
Only in special cases may one observe both self-assembled structures and phase coexistence (e.g. Figures \ref{fig:saPictures} and \ref{fig:vle} and Refs. \onlinecite{sciortino_phase_2009, munao_phase_2014, reinhardt_re-entrant_2011, mahynski_grafted_2015}).

The results in this study are consistent with the experimental and computational results of Wolters \textit{et. al.}, who studied a trimer fluid similar to the ones studied in this work.\cite{wolters_self-assembly_2015}
The results in this study may also be used to guide future experiments on tuning the shape of the trimers to control the formation of self-assembled structures that have not been observed in experiments.
While the attractive interaction in Wolters \textit{et. al.} is shorter-ranged than in this work, the fluid phase behavior of their model can be cast within the context of this work by matching the second virial coefficient, which assumes Noro-Frenkel extended corresponding-states applies.\cite{noro_extended_2000, foffi_possibility_2007}
Although we have not computed the phase diagram for the fluid studied by Wolters \textit{et. al.}, we can anticipate its phase behavior by comparing the second virial coefficient, $B_2$, with experiment, and also the Boyle temperature with Table \ref{tab:modelparams}.
This comparison is made based upon a trimer model, referred to as the MM-LJ model in this work, using the same shape parameters of Wolters \textit{et. al.}, $L=0.57\sigma$, $\Theta=91$ degrees and smaller repulsive bead sizes, $\sigma_r = 0.85 \sigma$, but with the potential described in Section \ref{sec:models} of this work.\cite{wolters_self-assembly_2015}
Wolters \textit{et. al.} reported a second virial coefficient of $B_2/\sigma^3 \approx -11$, which corresponds to a deplentant concentration of $\phi_d \approx 0.2$.
This value of $B_2/\sigma^3$ in turn corresponds to a reduced temperature of $k_B T/\epsilon = 0.355 \pm 0.005$ for the MM-LJ model.
The results in this study are consistent with the possibility that the MM-LJ model forms only elongated clusters, as found in the work of Wolters \textit{et. al.}, because the Boyle temperature for the MM-LJ model is $k_B T_\textrm{Boyle}/\epsilon = 0.735 \pm 0.005$. 
This value of the Boyle temperature lies between models in Table \ref{tab:modelparams} that both formed elongated clusters, but the models transitioned from forming spherical clusters at $L=0.75\sigma$ to not forming spherical clusters at $L=0.4\sigma$.
One possible conjecture is that tuning the trimer shape in experiments for increased repulsion with respect to attraction (e.g. increased size of repulsive beads, $\sigma_r$, and $L$) may lead to the formation of spherical clusters.

%***********************************************************
\section{\label{sec:conclusions}Conclusion}
%***********************************************************

The phase diagrams of trimer particles with one central attractive bead and two repulsive beads were computationally mapped out as a function of the trimer shape.
It has recently been shown that it is possible to synthesize similar colloidal trimer particles,\cite{kraft_surface_2012, wolters_self-assembly_2015} and this computational study may guide future experimental studies of different trimer geometries.
The trimer particles self-assembled into spherical clusters, elongated clusters and packed cylinders.
The shape of the trimers, and the state conditions, played a role in determining the type of self-assembled structure that is formed.
In addition, some trimer geometries led to macroscopic fluid phase separation.
The transition from microscopic self-assembly to macroscopic fluid phase separation may be understood in terms of the relative size of repulsion and attraction in the particle.
In special cases, both self-assembled structures and macroscopic phase separation occurred simultaneously.

While the effect of the shape of the trimer on the phase behavior is the emphasis of this study, future investigations may utilize interaction potentials that model a particular system more closely (e.g. shorter-ranged interactions for patchy colloids).
Note that the continuous potential in this work was chosen to make the model amenable to molecular dynamics simulations, which will be the subject of future publications to study the kinetics of assembly.

The obvious case of $L=0.5\sigma$ was omitted from this study for the following reasons.
Although extensive simulations were conducted for $L=0.5\sigma$, the state conditions where the fluid potentially exhibited phase separation and/or self-assembly involved low temperatures.
Our existing set of Monte Carlo moves were not sufficient to sample these conditions adequately.
Proper sampling under these conditions may require more sophisticated cluster trial moves (e.g. Refs. \onlinecite{whitelam_avoiding_2007, whitelam_role_2009, ruzicka_collective_2014}).
Additional study of the cases near $L/\sigma=0.4$ and $0.5$ may be the subject of future publications.

\begin{acknowledgments}
H.W.H. acknowledges support from a National Research Council postdoctoral research associateship at the National Institute of Standards and Technology.
J.M. acknowledges support of the National Science Foundation (CBET-1120399), and the high-performance computing capabilities of the Extreme Science and Engineering Discovery Environment (XSEDE), which is supported by the NSF (TG-MCB-120014).
\end{acknowledgments}

\appendix
%***********************************************************
\section{\label{app:saexample}Examples of Determining Self-Assembly Structural Transitions}
%***********************************************************

\begin{figure}
\begin{center}
\includegraphics[width=8.5cm]{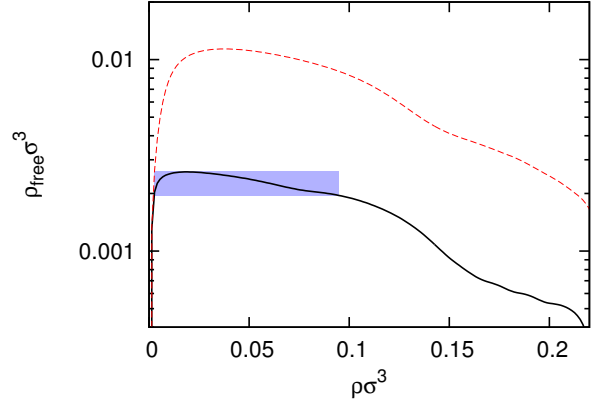}
\caption{
Number density of free trimers and premicellar aggregates, $\rho_{free}$ with (black solid line) $k_BT/\epsilon=0.25$ and (red dashed line) $k_BT/\epsilon=0.3$.
The blue shaded region shows where $\rho_{free}$ is within 75 \% of its maximum value.
$L=\sigma$, $\Theta=\pi/3$, and $V=729\sigma^3$.
}\label{fig:cmcdef}
\end{center}
\end{figure}

\begin{figure}
\begin{center}
\includegraphics[width=8.5cm]{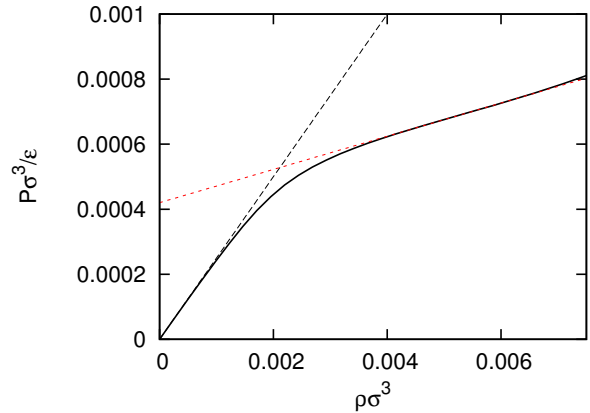}
\caption{
Pressure as a function of trimer density for $k_BT/\epsilon=0.25$ with the ideal pressure shown by the black dashed line and the fit to the second linear regime shown by the red dotted line.
$L=\sigma$, $\Theta=\pi/3$, and $V=4096\sigma^3$.
}\label{fig:idealdef}
\end{center}
\end{figure}

\begin{figure}
\begin{center}
\includegraphics[width=8.5cm]{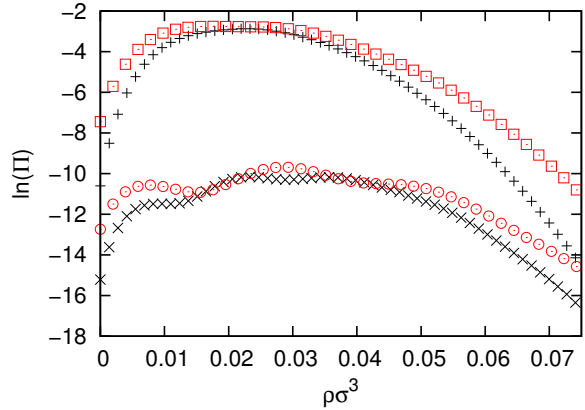}
\caption{
The probability to observe a number of trimers, $\Pi$ for $L=\sigma$, $\Theta=\pi/3$ with (black x)  $k_BT/\epsilon=0.275$, $V=729\sigma^3$ (red circle) $k_BT/\epsilon=0.275$, $V=512\sigma^3$ (black +) $k_BT/\epsilon=0.3$, $V=729\sigma^3$ and (red square) $k_BT/\epsilon=0.3$, $V=512\sigma^3$.
When $k_BT/\epsilon=0.275$, $\mu/k_BT = 3 ln(\sigma \Lambda) - 5$.
Otherwise, $\mu/k_BT = 3 ln(\sigma \Lambda) - 4.5$.
$\Lambda$ is the thermal de Broglie wavelength.
Probability distributions are shifted by a constant for clarity.
}\label{fig:cmtdef}
\end{center}
\end{figure}

\begin{figure}
\begin{center}
\includegraphics[width=8.5cm]{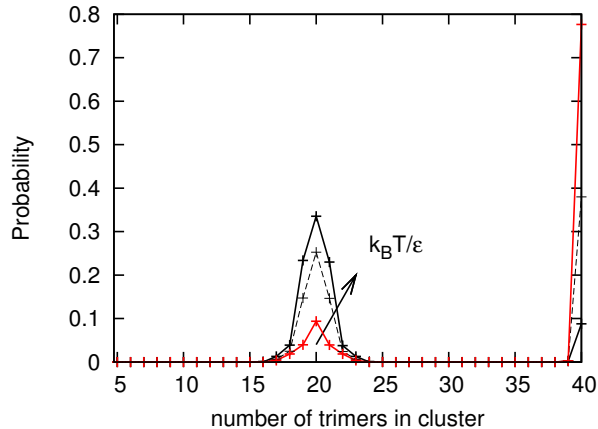}
\caption{
Probability distribution of number of trimers in a cluster in parallel tempering simulations for $L=\sigma$, $\Theta=\pi/3$, $V=729\sigma^3$, and $k_BT/\epsilon=0.154,0.16,0.167$.
}\label{fig:cddef}
\end{center}
\end{figure}

\begin{figure}
\begin{center}
\includegraphics[width=8.5cm]{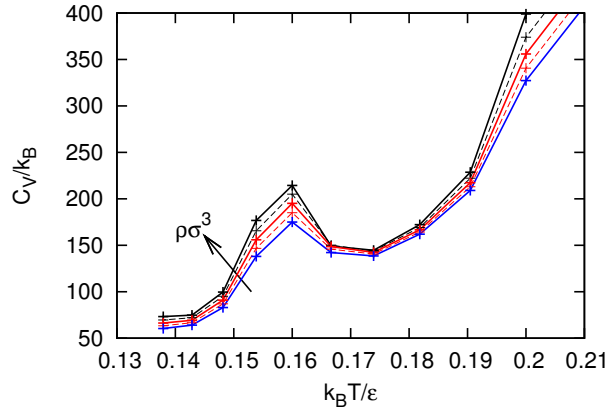}
\caption{
Grand canonical ensemble averaged constant volume heat capacity, $C_V$, in parallel tempering simulations for $V=729\sigma^3$, $L=\sigma$, $\Theta=\pi/3$, and N=35,36,37,38,39.
}\label{fig:cvdef}
\end{center}
\end{figure}

In this appendix, examples of the structural transitions from the spherical micellar fluid are provided for select state points for $L=\sigma, \Theta=\pi/3$.
As described in Section \ref{sec:methodspb}, these transitions include the critical micelle concentration (CMC), the critical micelle temperature (CMT), the spherical micelle to elongated cluster transition, and the high density boundary of the micellar fluid.
The transition between a free trimer fluid and micellar fluid is defined by the critical micelle concentration (CMC).
The CMC was obtained both structurally and thermodynamically, as described in Section \ref{sec:methodspb}.
Figures \ref{fig:cmcdef} and \ref{fig:idealdef} illustrate these different approaches.
In addition, there is a critical micelle temperature (CMT), above which a trimer fluid exists without micelles.
This CMT is not a true critical point, but it is a useful construct that suffers from some arbitrariness.
As demonstrated in Figure \ref{fig:cmtdef}, micelles formed at $k_BT/\epsilon=0.275$ due to the presence of a system-size dependent density of a second peak in the macrostate distribution.\cite{salaniwal_competing_2003}
But they did not form at the higher temperature of $k_BT/\epsilon=0.3$.
Therefore, the CMT, $T_\textrm{cm}$, is between these two temperatures, and reported as $k_BT_\textrm{cm}/\epsilon=0.2875 \pm 0.0125$ in Figure \ref{fig:saPhaseDiagrams}.
There is also a high density boundary for the micellar fluid, where micelles deform to improve packing, and eventually form different structures (e.g. cylinders shown in Figures \ref{fig:saPictures}g and \ref{fig:saPictures}h).
This high density boundary was defined approximately as the density at which the concentration of free trimers and premicellar aggregates is no longer constant.
Finally, at lower temperatures, there is a transition between spherical micelles and elongated clusters.
The structural definition of this transition was the temperature at which the system had an equal probability to form two micelles, or one elongated cluster.
This is shown in Figure \ref{fig:cddef} as two spherical clusters of 20 trimers in size combined at lower temperatures.
The thermodynamic definition of the micelle to elongated cluster transition was the temperature at which there was a peak in the constant volume heat capacity, shown in Figure \ref{fig:cvdef}.
For all cases, the thermodynamic and structural definition agrees within the error bars.
The determination of the value of the spherical to elongated cluster transition temperature, and the errorbars, from two bracketing isotherms, are analogous to the precedure for the CMT described above.

%***********************************************************
\section{\label{sec:exvol}Excluded Volume}
%***********************************************************

Excluded volume, $V_{ex}/\sigma^3$, was computed assuming hard sphere diameters of size $\sigma$.
In practice, the excluded volume was computed numerically by overlaying the trimer with a cubic grid of $N_{p}=10^9$ points and a side length, $V_{cube}^{1/3}$ equal to $\sigma$ plus the maximum intra-particle distance from a site to the center-of-mass.
By counting the number of grid points, $n_{o}$ which overlap with the trimer, $V_{ex}=\frac{n_{o} V_{cube}}{N_{p}}$.
By computing the excluded volume of a unit sphere and comparing to $4\pi/3$, the numerical error is expected to be on the order of $10^{-4}$.

%***********************************************************
\section{\label{sec:boyle}Second Virial Coefficient and the Boyle Temperature}
%***********************************************************

The second virial coefficient, $B_{2}(T)$, was calculated by Monte Carlo integration.
\begin{eqnarray}
B_{2}(T)=-0.5\int_V d\bm{r}f(\bm{r}) = -\frac{V}{2n}\sum_i^{N_{trial}}f(\bm{r}_i)
\\
f(\bm{r}) = e^{-U(\bm{r})/k_BT}-1
\end{eqnarray}
where $\bm{r_i}$ is the relative position and orientation of a second trimer with respect to the first trimer, and $i=1,...,N_{trial}$ randomly chosen positions and orientations of a second trimer with respect to the first.
In practice, the cubic volume, $V$ was chosen such that $V^{1/3}$ is greater than the twice the potential cut-off plus four times the maximum intra-particle distance from a site to the center-of-mass.
Convergence was reached when $|B_{2}|/\sigma_{\textrm{block}} < 10^{-2}$ or $\sigma_{\textrm{block}} < 10^{-2}$, where $\sigma_{\textrm{block}}$ is the standard deviation obtained from block averages of size $N_{trial} = 10^6$.
The Boyle temperature, $T_{\textrm{Boyle}}$, was found by starting at $k_BT/\epsilon=0.15$ and incrementing $T$ by $0.01$.
The reported $k_BT_{\textrm{Boyle}}/\epsilon$ in Table \ref{tab:modelparams} was the average of the two temperature increments nearest $B_{2}=0$, and the reported error was $\pm 0.005$ to span the entire temperature increment.

\nocite{*}
\bibliography{trimertrunc}% Produces the bibliography via BibTeX.
\end{document}